\documentclass[preprint,12pt]{elsarticle}

\usepackage{amssymb}
\usepackage[hyphens]{url}
\usepackage[hidelinks]{hyperref}
\hypersetup{breaklinks=true}
\usepackage{tabulary}
\usepackage{tabularx}

\usepackage{algpseudocode,algorithm,algorithmicx}

\newcommand*\Let[2]{\State #1 $\gets$ #2}

\journal{Journal of Parallel and Distributed Computing}

\begin{document}

\begin{frontmatter}



\title{Efficient and Secure Flash-based Gaming CAPTCHA}
\author[CTI,JOR]{Monther Aldwairi\corref{cor1}} 
\author[CTI]{Suaad Mohammed}
\author[CTI]{Megana Lakshmi Padmanabhan}

\address[CTI]{College of Technological Innovation, P.O. Box 144534, Zayed University, Abu
	Dhabi, United Arab Emirates}
\cortext[cor1]{E-mail: munzer@just.edu.jo; monther.aldwairi@zu.ac.ae}

\address[JOR]{ Computer and Information Technology Faculty, P.O. Box 3030, Jordan University of Science
	and Technology, Irbid 22110, Jordan}

\begin{abstract}
With the growth of connectivity to smart grids, new applications, and the changing interaction between customer and energy clouds, clouds are more vulnerable to denial-of-service attacks. Efficient detection methods are required to authenticate, detect and control attackers. Completely Automated Public Turing test to tell Computers and Humans Apart, CAPTCHA, is one efficient tool to thwart denial of service attacks. The server presents the user with a client puzzle to solve in order to gain access to the service or website. The puzzle should be hard enough for computers, but easy for humans to solve. Several methods have been suggested including the popular image-based, as well as video-based, and text-based CAPTCHAs. In this paper, we present a new Flash-based gaming CAPTCHA to differentiate bots from humans. We propose a drag and drop client puzzle where the user will play a simple game to answer a visual question. Our method turns out to be convenient, easy for users and challenging for bots. Additionally, it has gaming aspect, which makes it interesting to users of all age groups.
\end{abstract}




\end{frontmatter}


\section{\textbf{ Introduction}}
\label{Intro}

Smart grids and associated energy clouds are supposed to provide cheaper power management and easier peer to peer transactions \cite{Anoh:2018:VMM:3231053.3231096}. However, security has been a major concern specially when dealing with energy cloud infrastructure. Denial of Service (DoS) attacks have been identified as one of the most important potential threats to cloud-based energy management \cite{6809180}. Distributed Denial of Service (DDoS) was deemed high impact attack on smart grids that might result in communication failure preventing urgent signals that could put the smart gird at risk \cite{Komninos2014}. Asri et al. \cite{Asri2015} proved through realistic simulation that DDoS is a very dangerous threat to the smart grid ecosystems.  
\\
Denial of service against cloud of energy is usually perpetrated through Bots or automated scripts \cite{BAKER2018242}. Bots are one of the major concerns for many communication networks, because of their ability to send large number of requests to servers or clouds. In addition, they can be used for manipulation of online surveys, review and polls, including the results of elections \cite{AlDuwairi2015}. Therefore, several efforts have been made to identify human users from automated programs or bots. In 1950, the researcher Alan Turing hypothesized that machines could pose as humans. To prove that, he designed a test where an evaluator could tell human from machine based on simple questions. The evaluator had to evaluate the answers from both respondents and decide which is human. Failure to do so proved that machines could pose as human \cite{Aldwairi2017}.
\\

In 1997, Anderi Broder came up with the idea of using a “puzzle” to distinguish between bots and human users \cite{Shirali-Shahreza2007}. In 2000, Yahoo! approached Carnegie-Mellon University and described their chat-room problem, where bots would pose as human and get information from users \cite{HenryS.BairdAllisonL.Coates2003}. Luis et al. started setting some criteria for a test that can be used to tell computers and human apart. This is when the term CAPTCHA (Completely Automated Public Turing test to tell Computers and Human Apart) came about. For a test to be considered a CAPTCHA, the code and the data must be publicly available. The idea was to have a test that will prevent bots from gaining access to services even if all people know what the code is and how the CAPTCHA works \cite{VonAhn2004}. Besides preventing automated machines from registering to websites, CAPTCHA is effective in slowing DoS, spam email, and password brute force attacks \cite{Hasan2017}.
\\
The earliest puzzles from the 2000 CAPTCHA project started with using distorted texts. Later researchers have classified CAPTCHA methods into three classes: OCR-based, visual non-OCR-based and non-visual \cite{Shirali-Shahreza2010}. The OCR-based method uses the shortcomings of the Optical Character Recognition (OCR) system to make it harder for automatic computer programs to read the CAPTCHA. Visual non-OCR-based methods use computer systems deficiency in identifying types of objects within images. Non-visual use other forms of CAPTCHA such as audio.
\\
Another classification of CAPTCHA made by researchers was text-based, sound-based and image-based schemes \cite{Yan2008}. The text-based is the same as the earlier OCR-based method, where the text is distorted and rendered unreadable to bots. Sound-based schemes fit in the area of the non-visual method, and image-based scheme corresponds to the visual non-OCR method of the grouping done by Mohammad and Sajjad \cite{Shirali-Shahreza2010}. 
\\
Most modern CAPTCHAs are either solvable by machines, difficult for users, require technical resources, processing power or other requirements. In this paper, we proposed a new CAPTCHA technique: Flash-based gaming CAPTCHA, which requires fewer resources, easier for all users to solve, but still difficult for machines to pass. In order to evaluate the proposed CAPTCHA, the framework developed by Jakob Nielson \cite{JacobNielsen2012} was used. The learnability, efficiency, memorability, errors, and user satisfaction were measured. In addition, known attacks and technical requirements were added to the evaluation criteria. 
\\The rest of the paper is structured as follows. Section 2 surveys the related work. Section 3 describes the proposed method while the evaluation and results are described in section 4.

 \section{Related Work}
 \label{Related}
We follow the categorization by Mohammad and Sajjad \cite{Shirali-Shahreza2010} and present OCR-based, visual non-OCR-based and non-visual methods.

 \subsection{OCR-Based Methodsk}
\label{OCR}
OCR-based/text-based CAPTCHA has been used mostly by websites such as Google, Microsoft, TicketMaster, Yahoo, eBay, PayPal and many other major websites. Gimpy is one of the first CAPTCHAs developed by researchers at Carnegie Mellon University. The website implementing this method picks seven words from a dictionary and then slightly distorts them in a way that is readable by a human but not by machines. The test is then presented to the user who is asked to type in three of the words presented \cite{VonAhn2004} as shown in Figure 1.a. Ling-Zi and Yi-Chun \cite{Xiao2012} studied a login form of a popular Chinese bank website and pointed out vulnerabilities. They demonstrate how easy text-based CAPTCHA can be defeated with simple tools. They concluded that without appropriate design, CAPTCHAs implementation can be complicated and risky. The authors provided some key points for CAPTCHA designers to improve the security implementation of the text-based CAPTCHA and build a robust CAPTCHA scheme \cite{Switch}.
\\
Figure 1.b. shows another method called EZGimpy, where the program distorts the text by adding line, different colors and noises to the background to make it harder for machines to read \cite{Hajjdiab2017}. Mori and Malik \cite{Mori2003} developed a method where they identified three words of the Gimpy method with a 33\% success rate, and successfully solved the EZGimpy CAPTCHA 92\% of the time. 
\\
ScatterType \cite{Baird2005} in Figure 1.c. is another type of OCR-Based CAPTCHA, where the letters are scattered horizontally and vertically, making it harder for the machines to read. The legibility of the words depends on horizontal and vertical scatter distances, and the letters used. 
\\
Figure 1.d. shows Pessimal prints. It uses low-quality images that are synthetically generated by machines as a way to identify human user \cite{HenryS.BairdAllisonL.Coates2003}. The difficulty for OCR machines to read the words depends on the blur and threshold factors; the lower the numbers are the harder it is for machines to read the words. 

\begin{figure*}[!htbp]
	\centering 
	\includegraphics[width=\linewidth]{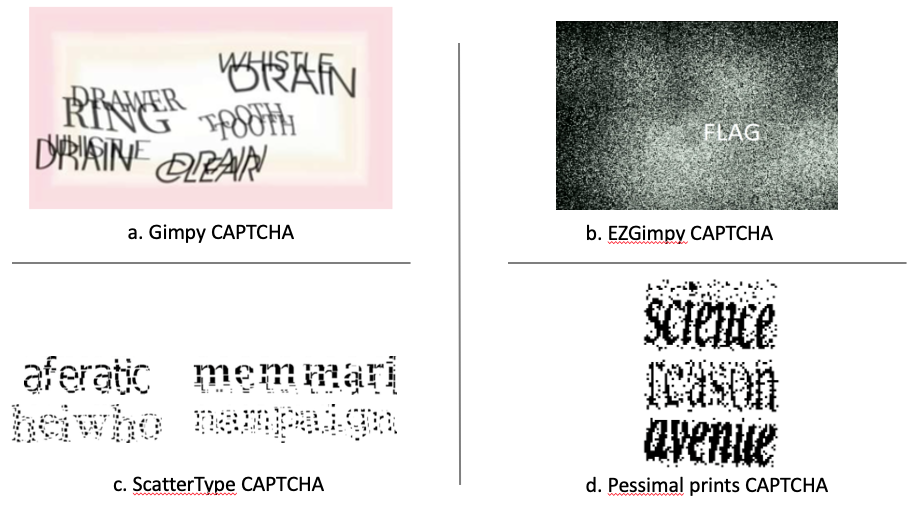}
	\caption{{Examples of OCR-Based CAPTCHAs}}
	\label{f1}
\end{figure*}

\subsection{Visual Non-OCR-Based Methods }
\label{NonOCR}

Difficulties associated with computer vision systems are used in these types of CAPTCHAs to identify and recognize objects in the image. Some examples of Visual non-OCR-based methods (image-based methods) include online collage and PIX. The ESP/PIX is among the first visual-non-OCR methods developed by Carnegie Mellon University. The program relies on a database that has a large number of concrete objects such as cats, flowers, cars, houses, horses etc. The program randomly picks an object and then randomly selects four images of that object. The images are then randomly distorted and presented to the users. From a drop-down menu, the users are expected to answer the question: “what are these pictures of?” \cite{10.1007/3-540-39200-9_18}. Figure 2.a below presents an example of ESP/PIX CAPTCHA. Kulluru et al. \cite{Suramwar2015} highlighted weaknesses of ESP Pix CAPTCHA. The first weakness is language dependency as it is available only English. Sometimes image recognition becomes awkward due to ambiguity present in object’s picture and it is also posing a big challenge low vision and visually impaired users. 
\\
Collage CAPTCHA is another example, where a bank of generic words is prepared. Such words include car, cat, fruit, flower, house, and airplane. The program randomly chooses a few words from the list of words and then searches the database for those pictures. The selected pictures are rotated a little and randomly located in the display space where they do not overlap. The computer then chooses a word and asks the user to select the responding picture. If the answer is correct, the computer assumes that the user is a human \cite{Shirali-Shahreza2007a}. Figure 2.b. shows an example of this method. Later Shanker et al. \cite{Shanker2013} proposed hybrid collage CAPTCHA to make it more difficult and resistant to hacker attacks. High usability and response time were achieved in hybrid college CAPTCHA as compared to collage CAPTCHA.

\begin{figure*}[!htbp]
	\centering 
	\includegraphics[width=\linewidth]{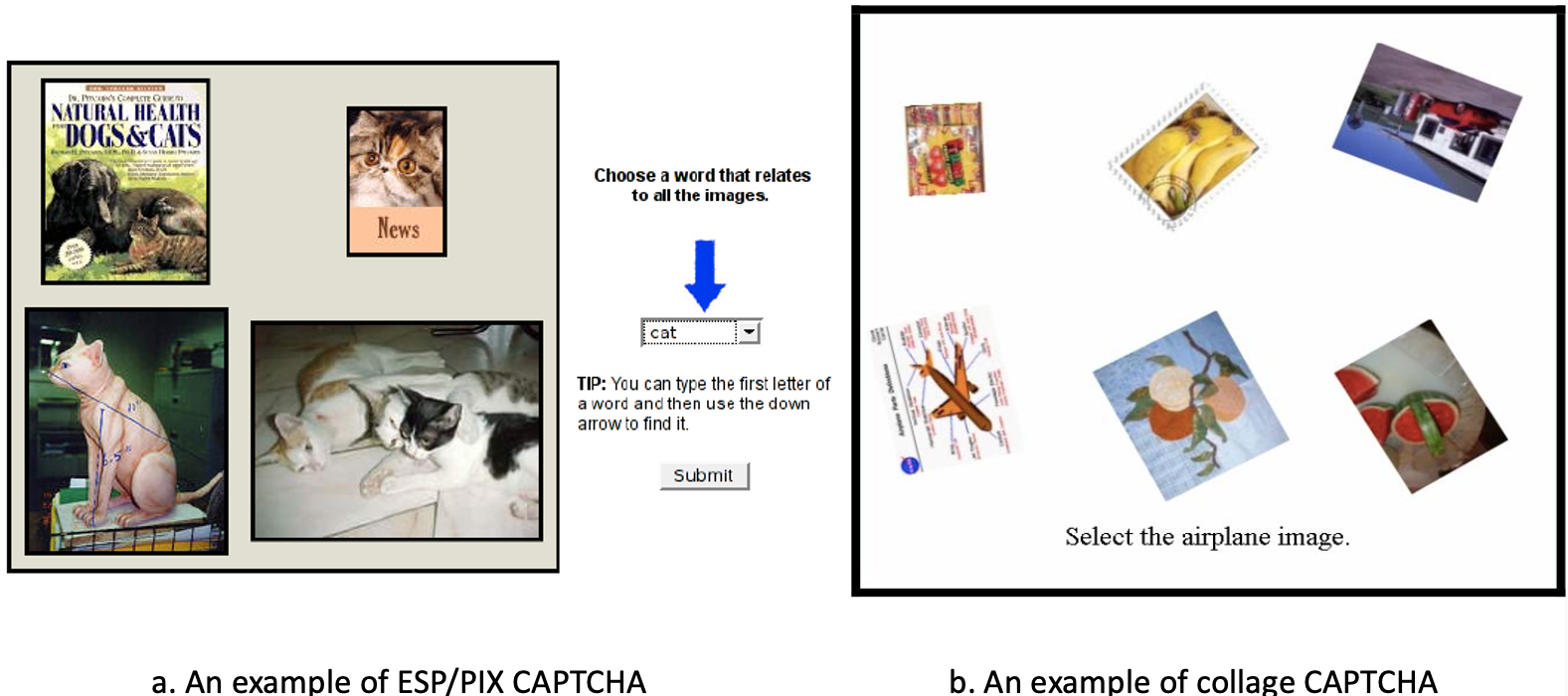}
	\caption{{Examples of Visual-non-OCR CAPTCHAs}}
	\label{f2}
\end{figure*}
Mohammad and Sajad implemented a “Drawing CAPTCHA” using a Java Applet inserted in a webpage that can run easily on any website to achieve security against DoS attacks. Asirra standing for Animal Species Image Recognition for Restricting Access \cite{elson2007asirra} was a common image-based CAPTCHA introduced by Microsoft in 2007. It displays 12 images for both cats and dogs, and then asks the user to identify cats’ images and select them. This task can be easily accomplished by human and may be challenging for machines.
\\
DeepCAPTCHA is another approach that has been proposed by Nejati et al. in 2014 \cite{Nejati2014}. The framework analyzes the 3D models of real-world objects, processes them so that machine cannot recognize the objects, and presents them to the user to sort based on their relative size. The authors believe machines cannot solve DeepCAPTCHA while user can easily solve it by using their reliable object recognition ability. The experimental results of their framework showed that individuals solved the DeepCAPTCHA with a high accuracy of 84\%.

\subsection{Non-Visual Methods }
\label{NonVisual}

Question-based CAPTCHAs are an example of non-visual CAPTCHA. These are methods where the user answers a specific question that needs reasoning \cite{Shirali-Shahreza2007}. Figure 3.a. presents an example of question-based CAPTCHA where the user needs to identify stationary objects and add their counts. Another example is math-question-based CAPTCHA, often called QRBGS CAPTCHA \cite{Hernandez-Castro2010}. Some CAPTCHAs that have been used are too easy while others are difficult for many users to solve such as the complicated math in Figure 3.b. These methods use very few numbers of patterns therefore, can be easily solved by a designed automatic solver \cite{Shirali-Shahreza2011}.
\begin{figure*}[!htbp]
	\centering 
	\includegraphics[width=\linewidth]{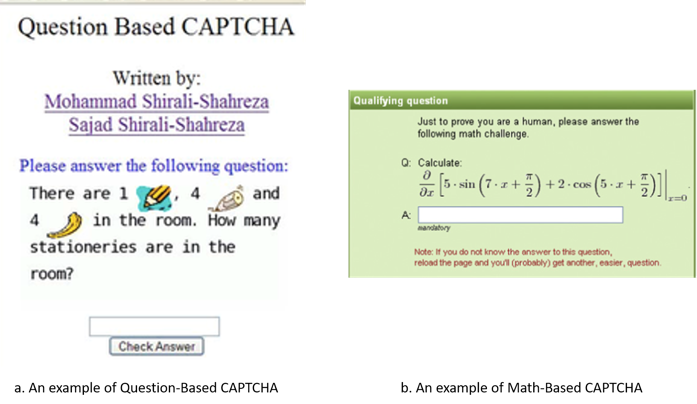}
	\caption{{Examples of Non-visual CAPTCHAs}}
	\label{f3}
\end{figure*}

\subsection{Energy Clouds Security}
\label{EngCloudSec}

While many researchers focused on the energy cloud optimization in terms of cost, stability, adaptability and sustainability, we believe security should not be overlooked \cite{BAKER201796}. With massive custom network-of-networks energy clouds proposed, such as Cloud-SEnergy's \cite{BAKER2018242}, that requires deployment of mission critical IoT devices, denial of service attack jumps to the forefront of the threat landscape \cite{su11133647}. While emerging research efforts focused on anomaly detection of DDoS in computing environments, CAPTCHA remains one of the most efficient low overhead techniques to curb DDoS campaigns \cite{Mahdavi_Hezavehi_2020}.

Due to security concerns about existing CAPTCHAs surveyed earlier, the area has gained a lot of researcher’s attention. Some suggested alternative text-based, images, while others suggested using audio and video techniques \cite{MohamedAlDhanhani2016}. However, these traditional methods are not safe anymore. For example, text-based CAPTCHA was the most used method in the last decade, but it is vulnerable to dictionary attacks, the existence of segmentation software and more \cite{AlRoum2018}. While audio-based CAPTCHA can be easily attacked by bots with voice recognition, and the ability to distinguish between the pronunciation of words and the volume of noise \cite{doi:10.1111/exsy.12404}. Video-based CAPTCHA can be easily attacked because of the ability of software to extract images from the video and analyze them to solve the challenge. Moreover, audio and video-based CAPTCHAs are not preferred by network applications because of their excessive consumption of bandwidth and time \cite{7894031}.

 \section{Flash-based Gaming CAPTCHA}
 \label{Flash}

We propose Flash-based gaming CAPTCHA as a way to tell users and machines apart. The idea is to present a simple and intuitive “game” for the user to play. A dataset of simple games requiring human experience will be designed and the website will randomly select a CAPTCHA game to present to the user. To solve this CAPTCHA, the user must identify the object and figure out the required action or question to be answered. Figure 4.a shows an example of the proposed CAPTCHA, in which user is expected to drag and drop the correct ball into the net. In this particular example the required object is the soccer ball and the action is to drag the ball into the goal net. After successfully solving the CAPTCHA, the user is allowed to access the required service. However, if the CAPTCHA is not solved correctly, the website displays an error message, and then the CAPTCHA is refreshed with a random one from the database. Figures 4.b and 4.c show examples of a successful and failed attempts, respectively.

\begin{figure*}[!htbp]
	\centering 
	\includegraphics[width=\linewidth]{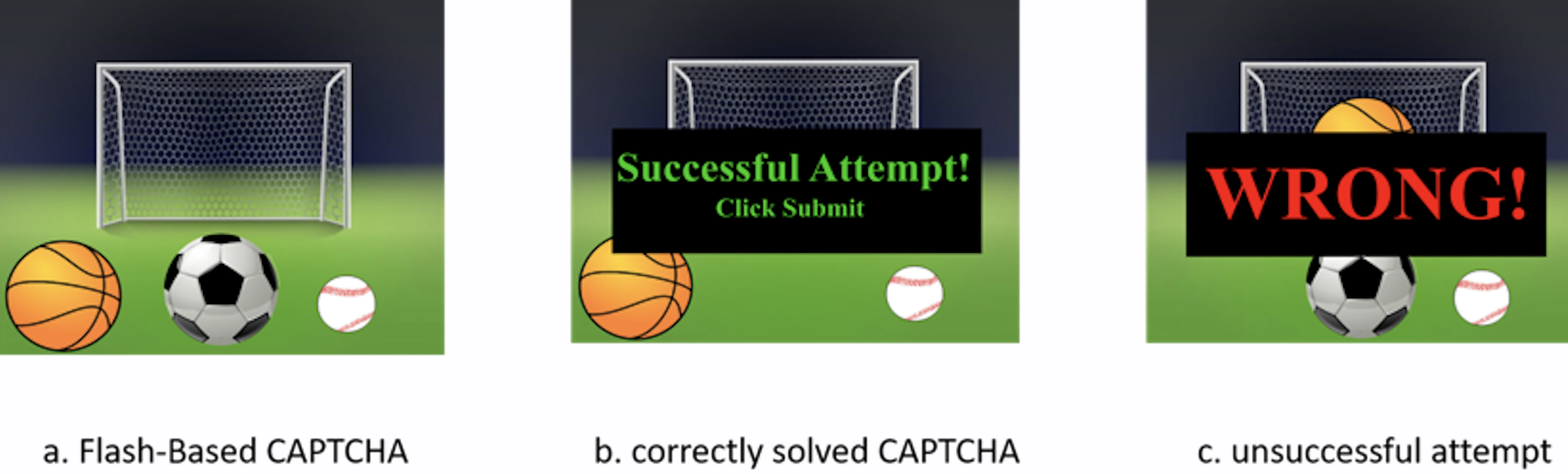}
	\caption{{Flash-Based Gaming CAPTCHA}}
	\label{f4}
\end{figure*}

The processing of proposed CAPTCHA is depicted by Algorithm 1. The user's mouse coordinates are collected on mouse press and release, and used to determine that he dragged the correct ball into the goal post. The proposed method is more secure and remains easy and fun for humans compared to traditional visual-based CAPTCHAs. It does not only require pattern or object recognition, but also commonsense knowledge and prior experience to perform the required action.

\begin{algorithm}
	\caption{Flash-based Gaming CAPTCHA
		\label{alg:CAPTCHA}}
	\hspace*{\algorithmicindent} \textbf{Input:} \textit{xy= mouse coordinates, zn= goalPostCoords} \\
	\hspace*{\algorithmicindent} \textbf{Output:} \textit{result}
	\begin{algorithmic}[1]
		\Statex
		\Function{VerifyGoal}{}
		\Let{($x,y$)}{$mouseCoordinates()$} 
				\State \Comment{ \% mouse click and hold signal is received\%}
		\If {$x,y \notin SoccerBallCoord$}
		\State $result\gets wrong$
		\Else
			\State \Comment{ \%mouse left release signal is received\%}
			\If {$x,y\in goalPostCoords$}
			\State $result\gets correct$
				\Else
					\State $result\gets wrong$
				\EndIf
		\EndIf
		\EndFunction
	\end{algorithmic}
\end{algorithm}

\section{Evaluation and Results}
 \label{Results}
 A user experience is carried out to evaluate the performance of the proposed CAPTCHA in terms of average time to solve, accuracy and ease of use by real users. We build a website to test Flash CAPTCHA implementation and capture the user’s reaction with a carefully designed survey. The survey asks the user to provide the following information before answering the CAPTCHA challenge.
 \begin{enumerate}
 	\item Age.
 	\item Gender.
 	\item Education level.
 	\item Years using the Internet.
 	\item Frequency of Internet use.
 	\item Vision problems.
 \end{enumerate}

 Online CAPTCHA was implemented and graduate students were surveyed after they finished solving the game. The survey data was collected using convenience sampling, which is based on a non-random sample. That is the results cannot be extrapolated to other than the participating sample. \\
 A total of 50 users participated in the experiment with 50\% male and 50\% female. The survey measures the users Internet surfing experience by asking about the years of using the Internet. Frequency of Internet use: daily, weekly or monthly. The results in Table 1, show that the average years of using the Internet for the 50 participants was 9.56 years with a range of 1-21 years, with the majority of the sample being aware of at least one type of CAPTCHA. Moreover, 68\% of the participants are using the Internet on a daily basis, 18\% of the participants are using Internet on a weekly basis and 14\% of the participants are using Internet on a monthly basis.
 
\begin{center}
	 \begin{table*}[!htbp]
		\caption{{Participants Internet Experience (Years)}}
		\label{participants}
		\ignorespaces 
		\centering 
	\begin{tabular}{ |c|c|c|c|c| } 

 		\hline 
			Years of Using the Internet &
			1-5 &
			6-10 &
			11-15 &
			16-21\\
			\hline
			Number of Participants &
			18 &
			13 &
			10 &
			9\\
			\hline 
	\end{tabular}
 \end{table*}
\end{center}

 The CAPTCHA code measures the average time required to solve the challenge. The time was measured using Action Script function \textit{getTimer()} as the difference of time between displaying the game and releasing the ball. Figure 5 shows the average time for users with varying Internet experience, the overall average time was 9.5 seconds. The figure shows that the average time required to solve CAPTCHA is not correlated to the users Internet usage and skills. The users who use the Internet on monthly basis were able to solve the challenge faster that users who use the Internet on weekly basis with 10.8 seconds. This proves that even users with low Internet and computing skills can easily understand and solve the challenge presented by the new CAPTCHA.
 
 \begin{figure*}[!htbp]
 	\centering 
 	\includegraphics[width=\linewidth]{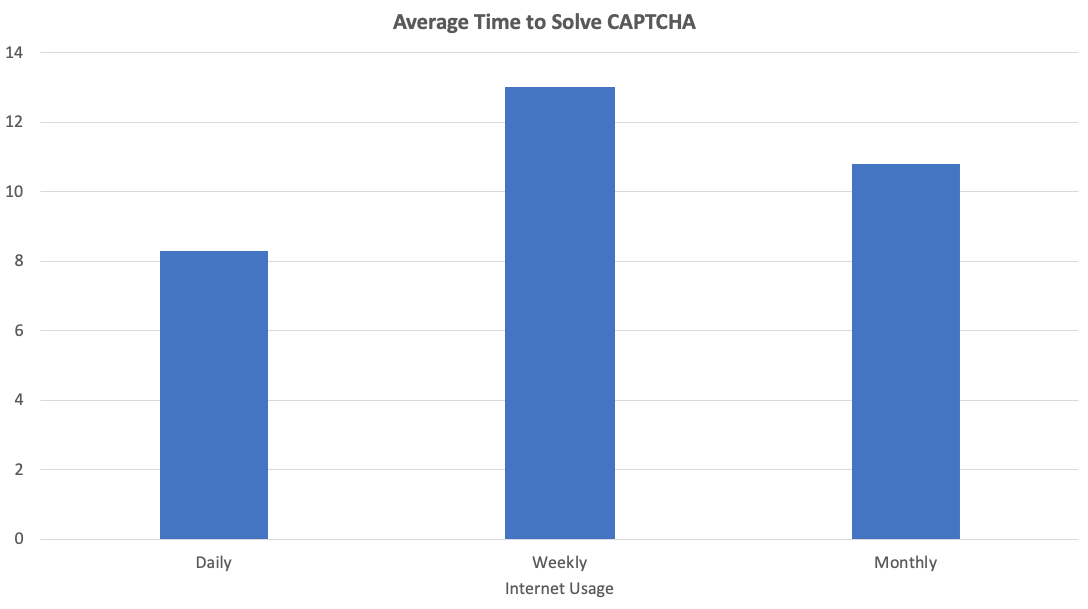}
 	\caption{{Average Time Required to Solve the CAPTCHA}}
 	\label{f5}
 \end{figure*}

 To be able to determine the ease of use of Flash-based CAPTCHA, participants of different ages and education levels were surveyed. The sample contains 50\% male and 50\% female, with ages ranging between 8 and 48, and the average age of 21.7 years. The participant’s education level varies as shown by Figure 6, where 22\% are primary level, 6\% are preparatory level, 16\% are secondary level, 8\% are iploma level, 32\% are Bachelor level, 14\% are Master level and 2\% had their Ph.D. This diverse sample was very important to identify how the age and different education levels may affect solving the CAPTCHA.
 
 \begin{figure*}[!htbp]
 	\centering 
 	\includegraphics[width=\linewidth]{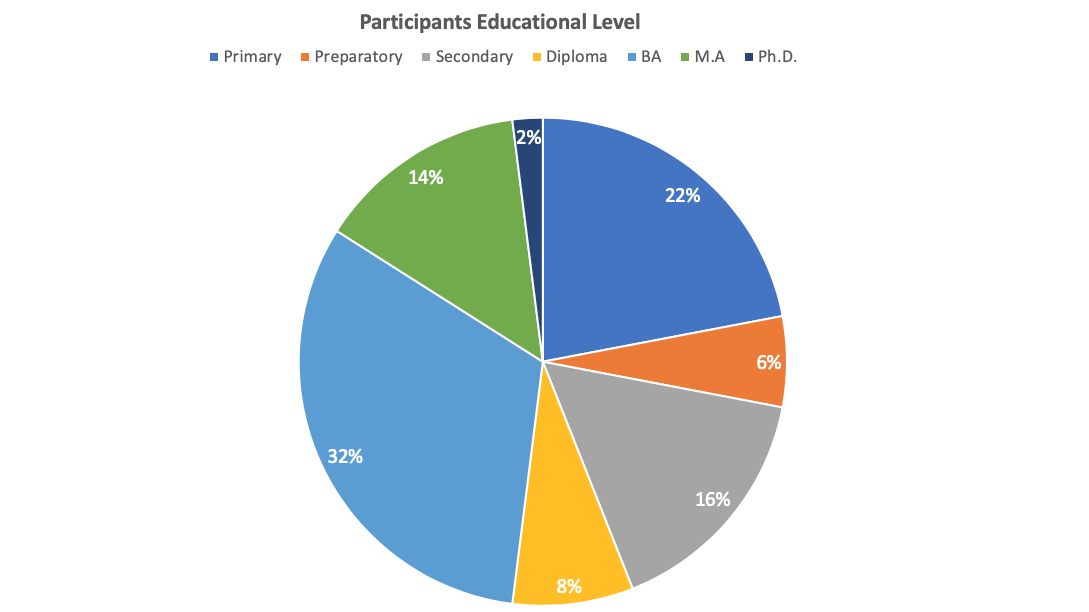}
 	\caption{{Participants Educational Level}}
 	\label{f6}
 \end{figure*}
 Furthermore, the participant’s vision issues were very important because we are using a Flash-based CAPTCHA. Seventy two percent of the participants did not have any vision impairments while 28\% have vision issues including nearsightedness and farsightedness. Figure 7 shows the number of failed attempts for each participants who had a vision impairments was 0, which means all of them were able to answer the CAPTCHA correctly. This supports that Flash-based CAPTCHA can be easily solved by users with vision impairment.
 
 \begin{figure*}[!htbp]
 	\centering 
 	\includegraphics[width=\linewidth]{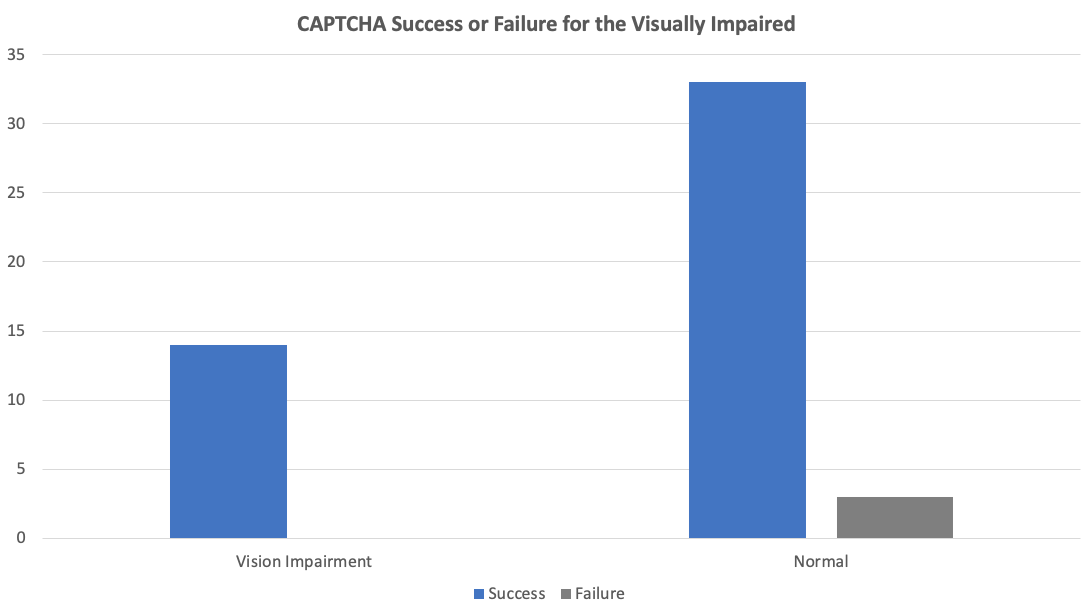}
 	\caption{{Visual Impairments Effects on Success and Failures}}
 	\label{f7}
 	
 \end{figure*}

\section{Comparison to Related Work}
\label{Comparison}

 The Flash-based CAPTCHA user preference was compared to six well-known CAPTCHAs in the literature including: Gimpy, ScatterType, Pessimal Prints, Collage, Math-based (QRBGS), and question-based.  In addition, we identify and discuss the weaknesses of each CAPTCHA in terms of technical requirements, easiness to bypass by bots and any known attacks against them. To do so, we use the framework developed by Jakob Nielson \cite{JacobNielsen2012}. The CAPTCHAs were analyzed, from users’ points of view, using the following metrics.

 \begin{enumerate}
 	\item \relax  Learnability: How easy is it for users to solve the CAPTCHA the first-time users are presented with it?
 	\item \relax  Efficiency: Once users are familiar with the design, how quickly they can solve the CAPTCHA?
 	\item \relax  Memorability: After not using this CAPTCHA for a while, how long will it take users to solve a similar CAPTCHA?
 	\item \relax  Satisfaction: How pleasant is it to use the CAPTCHA?
 	\item \relax  Errors: How many failed attempts did users make before solving the CAPTCHA?
 \end{enumerate}

 The survey used five-point Likert scale and answers ranging from 1 (very difficult), 2 (difficult), 3 (normal), to 4 (easy), and 5 (very easy). In order to make the analysis more convenient, we combine the scores 1 (very difficult) and 2 (difficult) together, 4 (easy) and 5 (very easy) together. In doing this, we obtain 2 charts comparing the level of easiness and the level of difficulty of the CAPTCHAs. The final results of the study for easiness are shown by Figure 8. The figure shows a summary of the most preferred CAPTCHAs according to the users.
 
  \begin{figure*}[!htbp]
 	\centering 
 	\includegraphics[width=\linewidth]{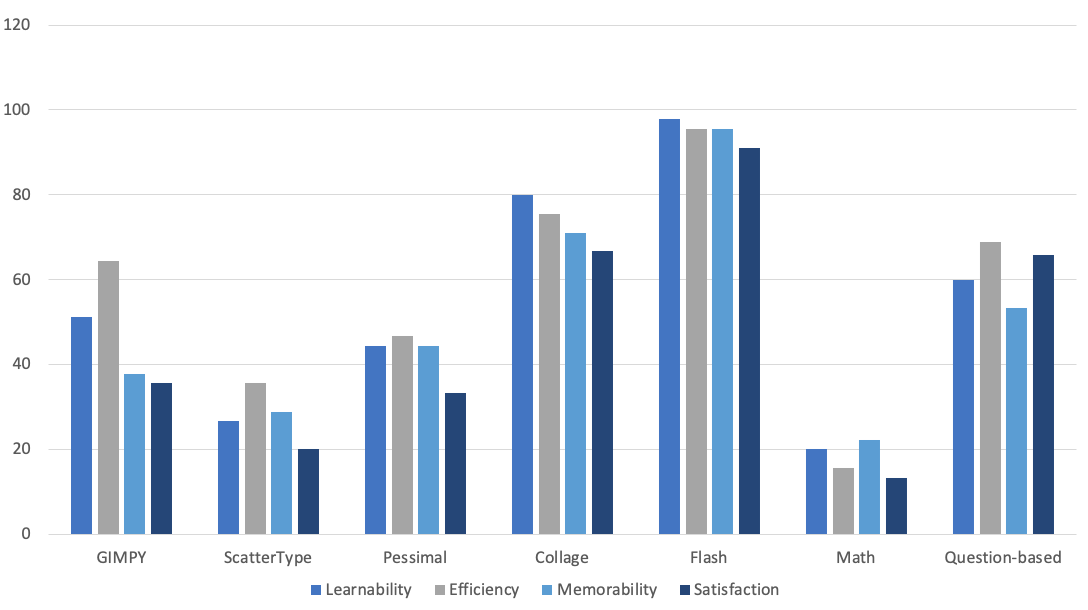}
 	\caption{{Users Preference}}
 	\label{f8}
 \end{figure*}

\subsection{Users Preference}
\label{Preference}

 According to the survey analysis, the Flash-based CAPTCHA is easiest and fastest to solve, easiest to remember and most pleasant to use. The subsections below explain how the users rated them according to their learnability, efficiency, memorability, and satisfaction.
 \subsubsection{Learnability}
 \label{Learnability1}

 With regards to learnability, 97.78\% of the respondents said that Flash-based CAPTCHA is the easiest to solve, 80\% of the respondents thought that Collage CAPTCHA is easy to solve, while the third easiest CAPTCHA to solve was the Question-based CAPTCHA with 60\% and then the Gimpy Captcha with 51.11\%. The Pessimal CAPTCHA, Scatter Type CAPTCHA and Math CAPTCHA scored of 44.44\%, 26.67\% and 20\%, respectively.
 
  \subsubsection{Efficiency}
 \label{Efficiency1}
 Taking the efficiency into consideration, 95.56\% of the respondents felt that the Flash-based CAPTCHA took the least time to solve, while 75.55\% of the respondents said that Collage CAPTCHA took less time to solve. Question-based, Gimpy, Pessimal, ScatterType CAPTCHAs came in next with  68.89\%, 64.44\%, 46.66\% and 35.55\%, respectively. Only 15.56\% of the respondents felt the Math-Based CAPTCHA took the least time to solve.
 
  \subsubsection{Memorability}
 \label{Memorability1}
 For memorability, 95.56\% of the respondents thought that Flash-based CAPTCHA is the easiest to remember and solve after a while of not using it, which is the highest percentage among all CAPTCHAs. The second in line with regards to memorability was the Collage CAPTCHA with 71.11\%, and then  Question-based CAPTCHA with 65.91\%. The Pessimal, GIMPY, ScatterType and Math-based CAPTCHAs came in last with 44.44\%, 37.78\%, 28.89\%, and 22.22\%, respectively.

 \subsubsection{Satisfaction}
 \label{Satisfaction1}
 A percentage of 91.11 of users felt the Flash-Based CAPTCHA was the most pleasant to use while 66.67\% said that Collage-based CAPTCHA was the most pleasant. Tthird in line was the Question-based CAPTCHA with 65.91\% followed by the Gimpy CAPTCHA with 35.56\%. Pessimal, ScatterType, and Math-based CAPTCHAs had 33.33\%, 20\% and 13.33\% respectively.
 
\subsection{Difficulty}
\label{Difficulty}
 We compare the seven CAPTCHAs in terms of difficulty to solve by the users. Figure 9 provides the user's opinions on the most inconvenient CAPTCHA with regards to learnability, efficiency, memorability and satisfaction.
 \begin{figure*}[!htbp]
 	\centering 
 	\includegraphics[width=\linewidth]{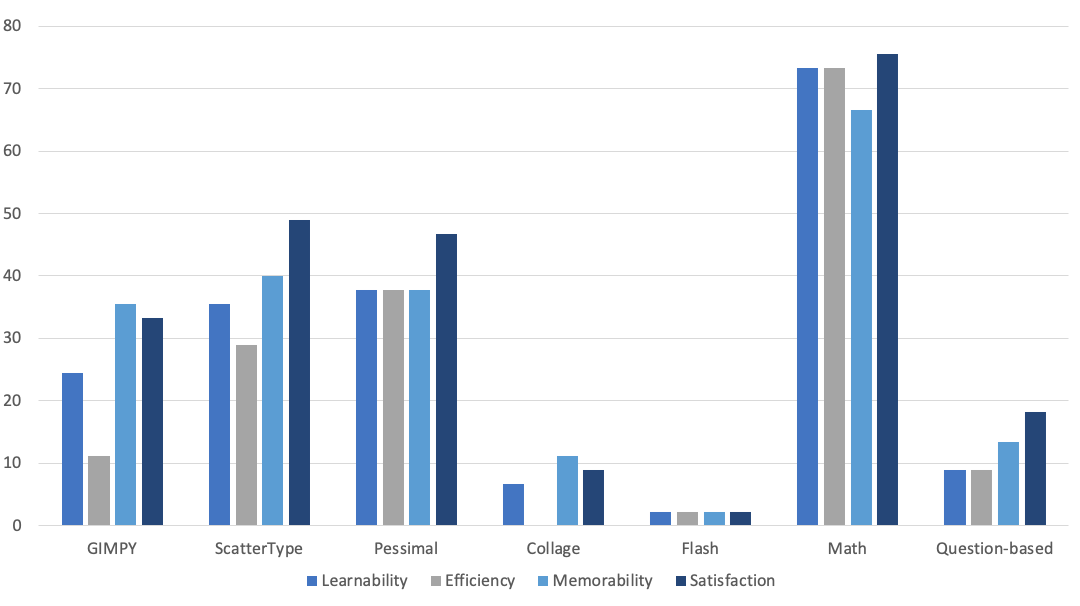}
 	\caption{{CAPTCHAs Difficulty}}
 	\label{f9}
 \end{figure*}

 \subsubsection{Learnability}
 Math-based CAPTCHA was the most difficult to use with 73.33\% of respondents, followed by Pessimal CAPTCHA with 37.78\% of the respondents finding it difficult to use. ScatterType, Gimpy, Question-based, Collage-based CAPTCHAs had the following percentages finding them difficult: 35.55\%, 24.44\%, 8.89\% and 6.6\%, respectively. Finally, only 2.22\% of the respondents found the Flash-based CAPTCHA to be difficult.
 
 \subsubsection{Efficiency}
 The Math-based CAPTCHA was the most inefficient with 73.33\% followed by the Pessimal with 37.78\%, ScatterType with 28.89\% and Gimpy with 11.11\% of the respondents taking more time to solve them. Question-based, Flash-based, and Collage-based CAPTCHAs took comparatively less time from the users with 8.89\%, 2.22\% and 0\%, respectively.
 
 \subsubsection{Memorability}
 Math-based CAPTCHA was the most difficult to remember how to solve for 66.67\% of participants after a long-time of non use. ScatterType CAPTCHA came in second with 40\% then Pessimal, Gimpy, Question-based and the Collage-based CAPTCHAs were found difficult to remember after a while by 37.78\%, 35.55\%, 13.33\% and 11.11\% of the respondents. Flash-based CAPTCHA was the difficult to remember only for 2.22\% of respondents.
 
\subsubsection{Satisfaction}
 Math-based CAPTCHA was found unpleasant by 75.56\% of the respondents and the ScatterType CAPTCHA was unpleasant by 48.89\% of the respondents. Pessimal, Gimpy, Question-based, and Collage-based CAPTCHAs were found unpleasant by 46.67\%, 33.33\%, 18.18\% and 8.89\% of the respondents, respectively. Only 2.22\% of the users found Flash-based CAPTCHA to be unpleasant.
 \\
 \subsection{Failure to Solve}
 According to the survey, the number of failed attempts varied from 1 to 64. Figure 10 shows the number of failed attempts given by the respondents for every CAPTCHA. The analysis shows that there have been at least 64 failed attempts to solve the Pessimal CAPTCHA, and at least 54 failed attempts to solve the ScatterType CAPTCHA. In the Pessimal CAPTCHA and ScatterType CAPTCHA the words are distorted and certain letters and numerals look alike such as ‘i’, ‘l’ and ‘1’... etc. Question-based and GIMPY CAPTCHAs had 35, and 34 failed attempts, respectively. Question-based CAPTCHA included pictures, which are small in size and difficult for the users to identify what is in the picture. At the same time, in a Gimpy CAPTCHA the texts are overwritten, which makes it disturbing for a user to view and type the text. Moreover, Six failed attempts to solve the collage CAPTCHA stemmed from unclear pictures. Finally, Flash-based CAPTCHA had the least number of wrong attempts of one. That is attributed to the fact the users find the CAPTCHA more interesting and clear. The math CAPTCHA is not considered in this analysis due to the fact that over 93\% of the respondents either did not answer the question or got it wrong. 
 
 \begin{figure*}[!htbp]
 	\centering 
 	\includegraphics[width=\linewidth]{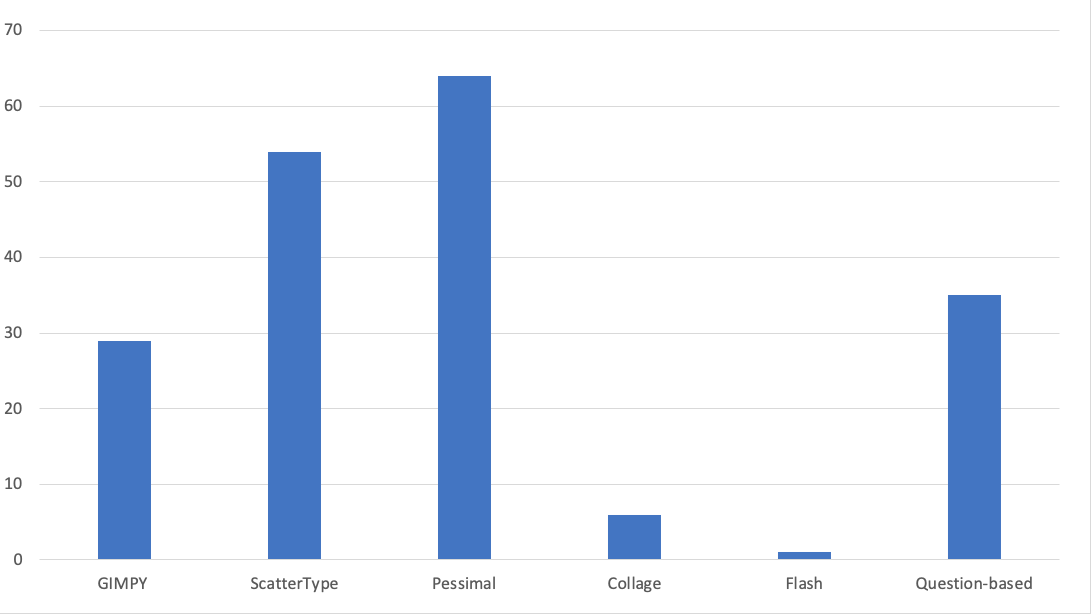}
 	\caption{{Number of Failed Attempts}}
 	\label{f10}
 \end{figure*}
 
 \subsection{technical requirement}
 The technical requirement is one of the main drawbacks for the CAPTCHAs. GIMPY, ScatterType, Pessimal, Collage, Question-based and the math-based CAPTCHAs require a bank of images, and a dictionary for these CAPTCHAs to work. However, Flash-based CAPTCHA requires a swf file per game, which is smaller in size than banks of images. In addition, it does not require a huge dictionary; therefore, the flash-based CAPTCHA requires fewer resources than the existing CAPTCHAs.
 \subsection{Resistance to Known Attacks}
Each one the aforementioned CAPTCHAs suffer from well-known attacks. An algorithm has been developed for passing a GIMPY CAPTCHA. The algorithm has a success rate of 33\% in identifying three words from the GIMPY CAPTCHA, rendering it insecure. 

The strength of Pessimal CAPTCHA against automated attacks depends on the blur and threshold factors; the lower the numbers the harder it is for machines to read the words \cite{HenryS.BairdAllisonL.Coates2003}. However, it was shown in the survey that this CAPTCHA had the highest number of wrong attempts and only 44.44\% learnability. Therefore, the CAPTCHAs that are used in websites need to have higher blur and threshold factors for them to be easily answered by a human, which will make it prone to attacks by machines. Therefore, Pessimal CAPTCHA is not the best CAPTCHA to be used to defend against automated attacks. 

There are some algorithms that can successfully solve advanced math problems and therefore bypass Math-based CAPTCHAs \cite{Hernandez-Castro2010}. However, math-based CAPTCHA is found to be the most difficult CAPTCHA for users to solve with only a success rate of less than 7\%. 

The readability of ScatterType CAPTCHA depends on the scatter distance: the lower the distance the more readable the CAPTCHA is. With a scatter distance of 0.15, only 27.78\% of the respondents stated that this CAPTCHA is easy. Therefore, the scatter distance has to be lower in order for a higher percentage of users to be able to read it. However, closer letter and lower scatter distance means that this CAPTCHA is subject to OCR attacks \cite{Baird2005}. 

Hacking Collage CAPTCHA depends on computer-vision recognition systems. When a machine comes across a picture it will search its own table to check if it came across the picture before. If yes the bot will look for the answer and if not it will save the picture into its system for later reference. A random guess can lead to about 17\% attack success rate. But the failure to solve remains high and makes it less than ideal CAPTCHA to deploy for users. 

For the Question-based CAPTCHA, a few wrong attempts are good enough for an automated machine to identify the answer for the respective CAPTCHAs. Bots save the pictures they detect into their database. When they come across the same image, it can review the previous responses given for that image.

Optical Character Recognition based attack does not work on a Flash-based CAPTCHA, this is because OCR is used by automated machines to identify characters and not images or Flash files. The CAPTCHA does not have a question and requires the cognitive thinking of a human to deduct what is required to be done. This makes it more resistant to known attacks.

\subsection{Complexity Analysis}
At first glance the presented client puzzle seems trivial, just put the proper ball into the goal net. This is simple for humans to do, but requires a Bot to have sophisticated levels of artificial intelligence and image processing. When it comes to brute forcing the game, we can easily make this simple scenario more complicated by adding different types of nets. Perhaps a score board would be introduced and the client is asked to even up the scores by dragging the ball to the proper team's goal net.
It is quite difficult to come up with a complexity analysis for this scenario, because playing a game (football) is much more complex than throwing a dice. 

 \section{Conclusions}
 \label{Conclusions}
 This paper proposed implemented and evaluated a new type of CAPTCHA: Flash-based CAPTCHA. According to the survey results, this CAPTCHA was the most convenient to use since it was voted the easiest to solve, with the least number of failures. In addition, it was found to be the fastest, the most pleasant to solve, and the easiest to remember after not using it for a long while. Moreover, Flash-based CAPTCHA needs fewer resources compared to the existing CAPTCHAs,  making it more efficient for use. Flash-based CAPTCHA is resistant to OCR attacks since this attack targets text-based CAPTCHAs, and the fact that this CAPTCHA needs cognitive abilities to solve, which makes it more resistant to automated attacks. Moreover, users from different age groups, levels of education, Internet skills and even those with vision impairments were able to solve it easily.
 
 \section*{Acknowledgment}
 This research was supported, in part, by Zayed University Research Office, Research Incentives Grant \# R18054.
 


 \section*{References}
  \bibliographystyle{elsarticle-num} 
  \bibliography{MyRef}


%\begin{thebibliography}{00}


%\bibitem{}

%\end{thebibliography}
\end{document}